\documentclass[twocolumn,showpacs,preprintnumbers,amsmath,amssymb]{revtex4}
\usepackage{graphics}
\usepackage{epsfig}
\usepackage{color}
\usepackage{subfigure}

\begin{document}

\title{Experimental demonstration of frequency pulling in single-pass 
free-electron lasers}
\author{E. Allaria$^{1}$, C. Spezzani$^{1}$, G. De Ninno$^{2,1}$}
\affiliation{1. Sincrotrone Trieste, S.S. 14 km 163.5, Basovizza (Ts), Italy \\
2. Physics Department, Nova Gorica University, Nova Gorica, Slovenia
}

\date{\today}

\begin{abstract}
Frequency pulling is a well-known phenomenon in standard laser physics, leading to a shift of the laser frequency when the cavity and maximum gain frequencies are detuned. In this letter we present the first experimental demonstration of frequency pulling in single-pass free-electron lasers. Measurements are performed using the single-pass free-electron laser installed on the Elettra storage ring. 
\end{abstract}

\maketitle

Frequency pulling is a well-known phenomenon in standard lasers, taking place when the peak of the gain spectrum is slightly detuned with respect to the frequency of one of the modes selected by the laser cavity. When this occurs, the lasing frequency is close to the one of the selected mode, but slightly ``pulled'' towards the maximum of the gain curve \cite{laser_books}. While generally investigated in the case of the continuous-wave operation of laser oscillators close to threshold, frequency-pulling has more recently become of interest also to other laser configurations \cite{fiber_pulling}. As an example, the importance of frequency pulling for a correct estimate of the laser-pulse parameters in the case of a mode-locked laser has been theoretically predicted and experimentally confirmed in \cite{modelock_pulling}. Moreover, frequency pulling has been exploited for the fine tuning of the laser frequency \cite{fir_pulling} and for beat-stabilization in multimode laser operations \cite{beat_pulling}.

In a free-electron lasers (FEL) \cite{book_FEL}, the light amplification is not due to the stimulated emission of an atomic system (like in standard lasers), but is instead relying on the  coherent emission of relativistic electrons oscillating into a periodic magnetic field generated by an undulator. 
In the early days of FEL's, the study of frequency pulling has been focused on (low-gain) oscillator systems \cite{book_FEL}. For such FEL configuration, the effect of frequency pulling is similar to that taking place in standard lasers:  the laser frequency is determined by the interplay between the proper frequencies of the cavity and that for which the gain is maximum \cite{cavity_fel_pulling, gain_fel_pulling}.
In the case of high-gain FEL's \cite{hfel_bonifacio}, the light amplification takes place during a single-pass of the electrons through undulators, without the use of a laser cavity. This motivates the strong interest in single-pass FEL's: being optics-free devices, they offer the unique opportunity to provide coherent and powerful emission in the spectral range from VUV to x-rays. In fact, the coherence of a single-pass FEL can be drastically improved if, before emitting, electrons are brought into interaction with an external coherent ``seed'' source (e.g., a laser).  As described more in detail in the following, such a seed determines the resonant frequencies at which the lasing effect occurs. Therefore, its action is somehow similar to the one of an optical cavity in standard lasers and oscillators FEL's. In a recent work \cite{epl_noi}, we have theoretically predicted that, also in seeded FEL's, a mismatch between the frequencies ``selected'' by the seed and the peak of the FEL gain curve may lead to a frequency-pulling effect. Based on numerical simulations, we proposed an empirical formula which generalizes the one commonly used for standard lasers and allows to properly describe the frequency pulling effect in a seeded FEL.   
In this work, we validate such a formula, providing the first experimental demonstration of frequency pulling in a seeded single-pass FEL.  Our results also aim at contributing to the lively debate carried out within the FEL community about the possibility to tune both the fundamental and harmonic wavelengths of a seeded FEL around those dictated by the seed. In this respect, we anticipate that, according to our results, the tunability that can be obtained through frequency pulling is quite limited, of modest interest for user experiments requiring a significant wavelength variation.       

In a seeded FEL, the process leading to coherent emission is induced by the interaction of the electron beam with an external optical pulse. When working in this configuration, the FEL may act as an amplifier and/or as an up-frequency converter of the input signal. In the first case, the seed pulse is typically characterized by a short wavelength, and by a relatively weak intensity. Effective seed sources of single-pass FEL amplifiers are, e.g., the high-order harmonics generated by the interaction of a high-power laser, typically a Ti:Sapphire, with a gas jet \cite{hhg}. As an alternative, the seed can be provided directly by the laser, or by one of its low-order (high-power) harmonics generated using, e.g., a BBO crystal. 
In this operation mode, called coherent harmonic generation (CHG), the FEL emission occurs at one of the harmonics of the input seed \cite{yu}.          

\begin{figure}
\centering
{\resizebox{0.46\textwidth}{!}{ \includegraphics{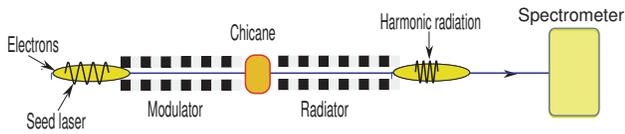}}}
\caption{Layout of the Elettra storage-ring FEL, including the diagnostics setup. For a detailed description of the system, see \cite{NIM}.}
\label{fig1}
\end{figure}

The results reported in this letter have been obtained on the seeded FEL installed on the Elettra storage ring, operated in CHG mode. The setup is shown in fig. \ref{fig1}. In the following, we briefly recall the principle on which CHG is based.  
The laser pulse is synchronized and overlapped to the electron bunch, while the latter travels through a first undulator (the modulator). The modulator is tuned to the same frequency $\nu_0$ and to the same polarization of the seed laser, thus satisfying a resonant condition that allows optimizing the energy transfer from the laser to the electrons. Such an energy transfer results in a periodic modulation of the longitudinal energy distribution. At the exit of the modulator, the electrons cross a magnetic chicane where the energy modulation is converted into 
a spatial modulation of the electron density. At the end of the chicane, the portion of the bunch that interacted with the seed laser presents a micro-bunched structure along the longitudinal axis, with Fourier components at the seed wavelength and at its higher harmonics \cite{prl_doppio}. When the micro-bunched electron beam travels trough a second undulator (the radiator), tuned to a harmonic of the seed $\nu_s=N_h\times \nu_0$ (where $N_h$ is an integer), the electrons radiate coherently and the produced intensity is proportional to the square of the number of involved electrons \cite{book_FEL}.
In \cite{epl_noi}, we have numerically demonstrated that the interplay between the harmonic frequency $\nu_s$ and the center frequency of the radiator gain curve, $\nu_u$, determines the FEL output frequency, $\nu_{CHG}$, according to the following formula:

\begin{equation}
\nu_{CHG} = \nu_s -(\nu_s -\nu_u ) \frac{\sigma _s ^2}{\sigma _s ^2 + \sigma _u ^2 }.
\label{fpul}
\end{equation}

Here, $\sigma _s$ and $\sigma _u$ are respectively, the FWHM widths of the seed and gain spectra. The frequency $\nu_u$ is given by 

\begin{equation}
 \nu_{u} = \left[\frac{\lambda_w}{2 c \gamma^2}\left(1+aw^2\right)\right]^{-1},
\label{eq:reso}
\end{equation}
where $\gamma$ is the electron-beam relativistic Lorentz factor,
$\lambda_w$ is the radiator period, $c$ the light speed and $aw$ is the undulator strength, which depends on the undulator gap.

In the following, we report about the first experimental demonstration of frequency pulling in a single-pass FEL and show that eq. (\ref{fpul}) provides a good description of the phenomenon. For the reported experiments, the Elettra storage ring was operated in single bunch mode, at a beam energy of 0.9 GeV. The seed wavelength was 391 nm, that is the second harmonic of the Ti:Sapphire laser. The measured laser energy per pulse was 0.9 mJ, and the bandwidth about 1.8 nm (FWHM). The radiator was tuned at 195 nm, i.e. is the second harmonic of the seed. As a consequence, $\sigma_s=1.8/2^2$ nm $=0.45$ nm.  

\begin{figure}
\centering
{\resizebox{0.48\textwidth}{!}{ \includegraphics{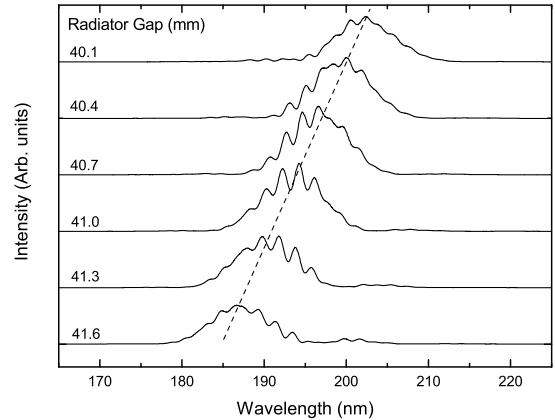}}}
\caption{Spectra of spontaneous emission as a function of the radiator gap. The straight line fits the positions of  the peaks of different spectra. The intensity oscillations in individual spectra are due to the interference between the radiator and modulator emissions. For these measurements, use has been made of the low-dispersion grating of the spectrometer.}
\label{fig2}
\end{figure}

\begin{figure}
\centering
{\resizebox{0.48\textwidth}{!}{ \includegraphics{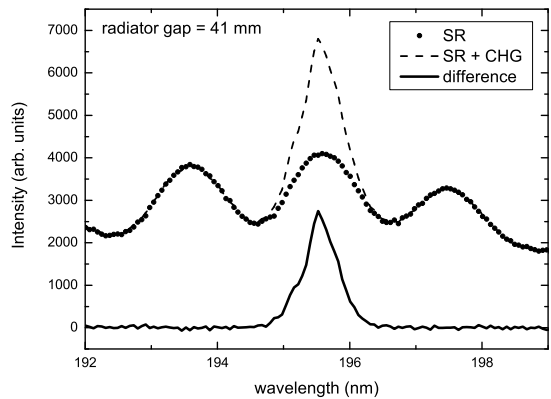}}}
\caption{Spectrum of the CHG and spontaneous (background) emissions (dashed line), of spontaneous emission only (dotted line) and of the CHG signal after background subtraction (continuous line). In order to better resolve the CHG spectrum, use has been made of the high-dispersion grating of the spectrometer.}
\label{fig3}
\end{figure}

In order to maximize the CHG signal \cite{prl2} and suppress the on-axis contribution at higher harmonics \cite{prl1}, the radiator was set in circular polarization. Also the strength of the dispersive section was optimized to maximize the CHG signal. The latter was analyzed by means of a UV spectrometer equipped with two different gratings for high and low dispersion, and a liquid nitrogen cooled  CCD detector (see fig. \ref{fig1}).

As a preliminary measurement, we have characterized the shift of the radiator gain curve, $\lambda_u=c/\nu_u$, as a function of the radiator gap. For that purpose, we have recorded the spectrum of the radiator spontaneous emission (i.e., no seed, no gain) for different gap values. The result is shown in fig. \ref{fig2}: a nearly linear shift of $\lambda_u$ is observed, corresponding to a variation of  about $-10.6$ nm/mm.

When CHG is active, the seeded electron bunch generates a light pulse which is four orders of magnitude more intense than the one generated by unseeded bunches \cite{prl2}. The repetition rate of the coherent signal is determined by the one of the seed laser, i.e. 1 kHz. The latter is more than one thousand times lower than the storage ring revolution frequency (1.16 MHz). Due to this, and to the quite long integration time of the CCD during spectrum acquisition (about 100 ms), a background subtraction is needed in order to get clean CHG spectra.  
An example of the procedure and of the resulting cleaned harmonic spectrum is shown in fig. \ref{fig3}. The measured bandwidth of the CHG emission is about 0.55 nm (FWHM). 

\begin{figure}
\centering
{\resizebox{0.48\textwidth}{!}{ \includegraphics{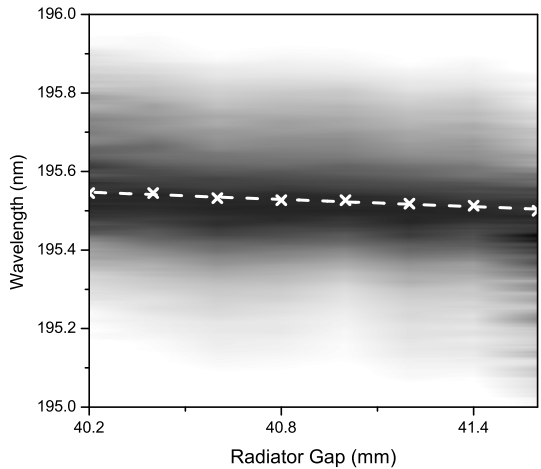}}}
\caption{CHG spectra (provided by cuts parallel to the vertical axis) as a function of the radiator gap. Darker areas correspond to spectra peaks. Peak wavelengths of different spectra are identified by the crosses. The line between peaks is a linear fit. }
\label{fig4}
\end{figure}

Finally, we measured the spectrum of the CHG signal for different values of the frequency $\nu_u$. As shown by eq. (\ref{eq:reso}), this can be done by varying $aw$, i.e. the radiator gap. The CHG spectra collected for different radiator gaps are shown in fig. \ref{fig4}. Since the efficiency of the harmonic generation process is reduced as the radiator is detuned with respect to the central wavelength, the data have been normalized to the maximum intensity (obtained at perfect tuning). Even if small, a shift of the FEL wavelength as a function of the radiator gap is clearly visible. A linear interpolation of the maxima (identified by the the crosses in the picture) gives a variation of 0.03 nm for a gap change of 1 mm. Using the calibration factor extracted from fig. \ref{fig2}, this corresponds to $|\Delta \lambda_{CHG}/ \Delta \lambda_u| \simeq 0.0028$, where $\lambda_{CHG}=c/\nu_{CHG}$. Such a value can be compared to the one provided by eq. (\ref{fpul}). A direct inspection of fig. \ref{fig2} gives $\sigma_u \simeq 8.6$ nm. Thus one gets:  

\begin{equation}
\frac{\Delta \lambda_{CHG}}{\Delta \lambda_u}=\frac{\Delta \nu_{CHG}}{\Delta \nu_u}=\frac{\sigma _s ^2}{\sigma _s ^2 + \sigma _u ^2 }\simeq 0.0027,
\end{equation}
which is in quite good agreement with the experimental result.

In this letter we provided the first experimental demonstration of frequency pulling in a single-pass FEL. The actual FEL frequency when the center of the gain curve is not superposed to the one of the seed spectrum is well predicted by eq. (\ref{fpul}). Our results thus confirm the predictions reported in \cite{epl_noi}. The measured effect is evident, although too small to be exploited in experiments requiring a significant tunability of the working frequency.

\end{document}